\documentclass[twocolumn,prd,amsmath,preprintnumbers,amssymb,superscriptaddress,nofootinbib]{revtex4}
\usepackage{hyperref}
\usepackage{graphicx}
\newcommand {\beq}{\begin{equation}}
\newcommand {\eeq}{\end{equation}}
\newcommand {\beqa}{\begin{eqnarray}}
\newcommand {\eeqa}{\end{eqnarray}}
\newcommand {\n}{\nonumber \\}

\newcommand {\p}{\partial}
\newcommand {\tr}{{\rm tr}}
\newcommand {\Tr}{{\rm Tr}}

\begin{document}

\preprint{KEK-TH-1293}

\title{String coupling and interactions in type IIB matrix model}

\date{\today}

\author{Yoshihisa Kitazawa}
\email{kitazawa@post.kek.jp}
\affiliation{High Energy Accelerator Research Organization (KEK), 
Tsukuba, Ibaraki 305-0801, Japan}
\affiliation{Department of Particle and Nuclear Physics
The Graduate University for Advanced Studies
Tsukuba Ibaraki 305-0801 Japan}
\author{Satoshi Nagaoka}
\email{nagaoka@post.kek.jp}
\affiliation{High Energy Accelerator Research Organization (KEK),
Tsukuba, Ibaraki 305-0801, Japan}

\begin{abstract}
We investigate the interactions of closed strings in IIB matrix model. 
The basic interaction of the closed superstring 
is realized by the recombination of two intersecting strings. 
Such interaction is investigated in IIB matrix model via two dimensional 
noncommutative gauge theory in the IR limit.
By estimating the probability of the recombination, we identify 
the string coupling $g_s$ in IIB matrix model.
We confirm that our identification is consistent with matrix string theory.
\end{abstract}

\maketitle
\tableofcontents

\section{Introduction}
\setcounter{equation}{0}

IIB matrix model \cite{IKKT} is considered as a candidate of the 
nonperturbative formulation of superstring theory.
The relation between IIB matrix model and perturbative string theory is
shown in \cite{KN3}.
Perturbative string theory is also contained in
Dijkgraaf, Verlinde and
Verlinde's matrix string theory \cite{DVV}.
Strong coupling region of two dimensional supersymmetric 
Yang-Mills theory is described
by the perturbative superstring theory. In the strong coupling limit,
a free Green-Schwarz string theory is obtained.
On the other hand,
weak coupling region is described by the perturbative Yang-Mills theory.
In addition, there is an intermediate region which is described by type IIB
supergravity solution in the large $N$ limit \cite{IMSY}.

The aim of this paper is to identify the string
coupling $g_s$ in IIB matrix model.
The hint for the identification comes from the matrix string theory.
The gauge coupling of the two dimensional Yang-Mills theory
has the
$[length]^{-1}$ of the worldsheet and it is related to the string
coupling as $g_{YM}^{-2} =\alpha' g_s^2$. 
The $g_s$ has the dimension $[length]$ on the worldsheet.
Thus, our task is to search for the dimensionful parameter on the
worldsheet.

Before searching for $g_s$,
we have to construct the worldsheets in IIB matrix model.
They are constructed as two dimensional classical backgrounds
\cite{KN3} in the IR limit. 
The string length is identified there and 
free multiple closed strings are obtained.
Vertex operators of type IIA superstring are
constructed from IIB matrix model on these backgrounds in
\cite{KN4}.
The relation between type IIA superstring theory and IIB matrix model is
also proposed in \cite{KS} in a different way.

In this paper,
we consider the interaction of perturbative strings.
The basic interactions are the transitions from 
two strings into one string or vice versa.
These interactions are introduced
in the formulation of superstring field theory in the light-cone gauge 
\cite{GSB}.
2 strings $\to$ 1 string interactions 
are represented by the recombination of 
intersecting strings locally. See figure 1.

\begin{figure}[hbt]
\begin{center}
\includegraphics[height=2cm]{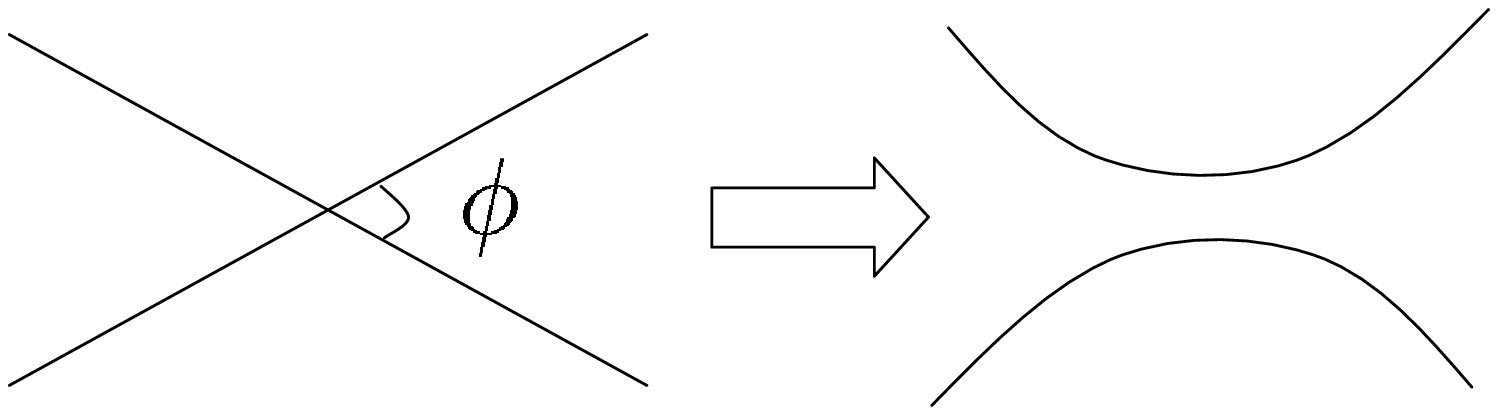} 
\caption{Two strings intersect at an angle $\phi$.
The recombination happens at the intersection point.
}
\end{center}
\end{figure}

Although this process locally represents the recombination of 
2 strings $\to$ 2 strings, the final state is connected globally.
Thus,
after the recombination, we obtain a single closed string.
In the matrix model, the instability of this system
comes from the off-diagonal modes. 
The system with the larger
intersection angle $\phi$ is more unstable than that with the smaller
intersection angle.
In other words, the decay rate of two closed strings is decided by the
intersection angle $\phi$.
We can replace the angle $\phi$ with another parameter
$q$ which is related to $\phi$
as $\phi =2 \tan^{-1} q$.
Since we choose the horizontal axis in figure 1 as the coordinate of the
worldsheet $\sigma$ and
the vertical axis as the
transverse scalar field $\phi_i (\sigma)$,
the parameter $q$ denotes the slope 
of the tilted strings.
In the two dimensional gauge theory, $q$ is 
the dimensionful parameter.
We clarify the $q$ dependence of the instability.
Since the configuration decays into the single closed string through the 
recombination process, we estimate $q$ dependence of 
the recombination probability.
In the string perturbation theory,
the recombination probability is proportional to
$g_s^2$ to the leading order.

In our investigation, we derive the action of multiple strings from
IIB matrix model.
We identify the unstable mode for the intersecting strings which
are the solutions of this action.
From this mode, we estimate the probability of the recombination.
By comparing the result to that of the perturbative string theory,
we identify the string coupling $g_s$ in IIB matrix model.

In section 2, we derive the action of multiple closed superstrings.
In section 3.1, we carry out the fluctuation analysis for 
the intersecting closed strings. We identify the unstable mode.
In section 3.2, we calculate the probability of the recombination from 
the unstable mode. By comparing our result with that of 
the perturbative superstring
theory, we identify the 
string coupling $g_s$.
In section 3.3, 
we estimate the probability of the recombination
in matrix string theory.
Section 4 is devoted to the conclusion.
In the appendix A, our notation of light-cone coordinates is given.
In the appendix B.1, the differential equations for the fluctuation modes are
solved. In the appendix B.2, 
the Schrodinger equation which controls the
time evolution of the probability is solved.

\section{The effective action of multiple strings}
\setcounter{equation}{0}

The interaction among multiple strings is described in 
various ways in string theory. 
Perturbatively, 
transitions from 
two strings into one string or vice versa
are the basic process.
We aim to identify such interactions in IIB matrix model.
Since this interaction is proportional to $g_s$ at the tree level,
we can identify $g_s$ in IIB matrix model through this process.

Let us start from the action 
\begin{eqnarray} \label{action}
S=-\frac{1}{g^2} \Tr
 \left(\frac{1}{4}[A^\mu,A^\nu][A_\mu,A_\nu]+\frac{1}{2}\bar{\psi} \Gamma^\mu
 [A_\mu,\psi] \right) \ , \hspace*{0.3cm}
\end{eqnarray}
where
$\psi$ is a ten dimensional Majorana-Weyl spinor and $A_\mu (\mu=0,
1,\cdots,9)$ and $\psi$ are $N \times N$ Hermitian matrices. 
By expanding the action (\ref{action}) around a two dimensional noncommutative
background, 
\begin{eqnarray}
[p_\mu,p_\nu]=i\theta_{\mu\nu} \ ,
\end{eqnarray}
we obtain an ${\cal N}=8$
two dimensional $U(n)$ noncommutative gauge theory \cite{CDS,AIIKKT,Li}
\begin{eqnarray}
S=-\frac{\theta}{8 \pi g^2} \int d^2 x \tr \left(
F^2_{\tilde{\mu}\tilde{\nu}} +2 (D_{\tilde{\mu}} \phi_i)^2
+[\phi_i,\phi_j][\phi_i,\phi_j] \right. \n
\left. 
+2 \bar{\psi} \Gamma^{\tilde{\mu}}
D_{\tilde{\mu}} \psi +2 \bar{\psi} \Gamma_i [\phi_i,\psi] \right)_* \ ,
\hspace*{1cm}
\end{eqnarray}
where $\tilde{\mu},\tilde{\nu}=0,1$ and $i,j=2,\cdots ,9$.

The $*$ product is described by
\begin{eqnarray}
a * b =\exp \left( \frac{iC^{\mu\nu}}{2} \frac{\p^2}{ \p \xi^\mu \p
		\eta^\nu} \right) a(x+\xi) b(x+\eta) |_{\xi=\eta=0} \ , \n
\end{eqnarray}
where $C^{\mu\nu}$ is defined as the inverse of $\theta_{\mu\nu}$.

By taking the commutative (IR) limit, we obtain a commutative gauge theory
\begin{eqnarray}
S=-\frac{\theta}{8 \pi g^2} \int d^2 x \tr \left(
F^2_{\tilde{\mu}\tilde{\nu}} +2 (D_{\tilde{\mu}} \phi_i)^2
+[\phi_i,\phi_j][\phi_i,\phi_j] \right. \n
\left.
+2 \bar{\psi} \Gamma^{\tilde{\mu}}
D_{\tilde{\mu}} \psi +2 \bar{\psi} \Gamma_i [\phi_i,\psi] +O(\theta)
\right) \ . \hspace*{1cm}
 \label{2d-YM}
\end{eqnarray}
Diagonal components are relevant degrees of freedom in the IR limit.
We interpret the diagonal elements of the field $\phi_i$ in (\ref{2d-YM})
as the coordinates of the fundamental strings.
This two dimensional Yang-Mills theory is related to a 
low energy effective theory of D-strings by the S-duality 
transformation, as is shown in fig. 2 of our previous paper
\cite{KN3}.

We map the worldsheet coordinate from $R^2$ into $R^1 \times S^1$
as
\begin{eqnarray}
z \equiv x_0+ix_1=e^{\tau+i\sigma} \ .
\label{r2s1}
\end{eqnarray}
Since the only gauge invariant quantity is a set of the eigenvalues
of the matrices $\phi_i$, if we go around the circle $S^1$, 
the eigenvalues can be interchanged. 
See figure 2.
We can consider a string of length $n$
by the identification $\phi_i (\sigma)=\phi_i (
\sigma +2 \pi n )$.

\begin{figure}[hbt]
\begin{center}
\includegraphics[height=4cm]{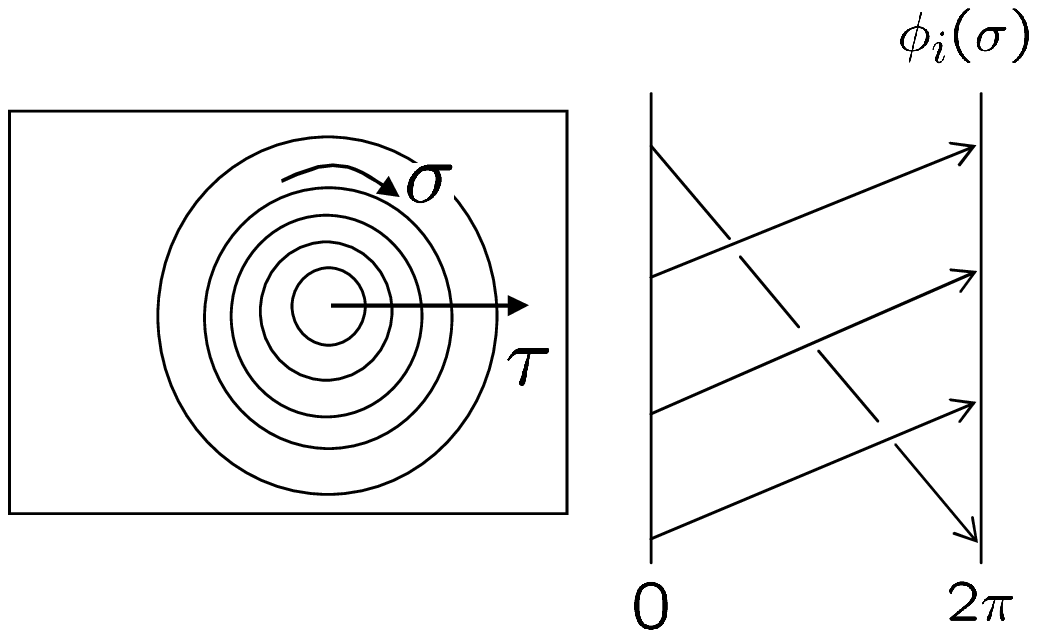} 
\caption{The string worldsheet coordinate.
We map the coordinate from $(x_0,x_1)$ into $(\tau,\sigma)$.
}
\end{center}
\end{figure}

Since multiple free closed 
strings can be regarded as a moduli space of this theory,
one can consider a configuration with multiple strings of various length
in general
\begin{eqnarray}
\phi_i (\sigma^a)=\phi_i (\sigma^a + 2 \pi w^a ) \ ,
\end{eqnarray}
where $n=\sum_{a=1}^{k} w^a$. 

For our purpose,
it is enough to analyze 
$U(2)$ gauge theory
since the recombination is a local problem which involves two strings.
At the tree level, the amplitude of this interaction is proportional to $g_s$.
After the coordinate transformation (\ref{r2s1}),
the action (\ref{2d-YM}) is mapped into
\begin{eqnarray}
S&=&-\frac{\theta}{8 \pi g^2} \int  
d\tau d \sigma 
|z|^2
\n &&
\tr \left(
2 \left(
\frac{\p_+ A_-}{z} -\frac{\p_- A_+}{\bar{z}}- [A_+,A_-]
\right)^2 
\right. \n && 
+4 \left(
\frac{\p_+ \phi_i}{z}- [A_+,\phi_i]
\right)^2
+ [\phi_i,\phi_j][\phi_i,\phi_j] 
\n &&
\left.
+2 
\bar{\psi} \Gamma^+ \left(
\frac{\p_+ \psi}{z}- [A_+, \psi]
\right) 
\right.
\n &&
\left.
+
2 
\bar{\psi} \Gamma^- \left(
\frac{\p_- \psi}{ \bar{z}}- [A_-, \psi]
\right)
+2  \bar{\psi} \Gamma_i [\phi_i,\psi] 
\right) \ .
\end{eqnarray}

\begin{figure}[hbt]
\begin{center}
\includegraphics[height=2.4cm]{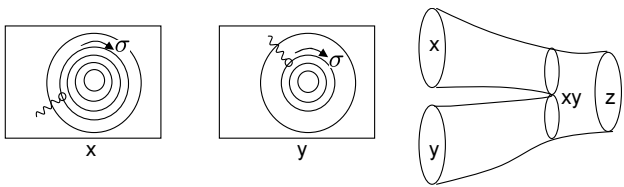} 
\caption{Two closed strings $x$ and $y$ recombine at some time 
and finally we obtain a single closed string $z$. 
}
\end{center}
\end{figure}

After the field redefinition :
\begin{eqnarray}
&&A_+ \to \frac{1}{z} \sqrt{\frac{g^2}{\theta}} A_+ \ , \quad
A_- \to \frac{1}{\bar{z}} \sqrt{\frac{g^2}{\theta}} A_- \ , \n
&&\psi_R \to \frac{1}{\sqrt{z}} \psi_R, \quad
\psi_L \to \frac{1}{\sqrt{\bar{z}}} \psi_L \ , \label{redef}
\end{eqnarray}
and the rescaling :
\begin{eqnarray}
\tau \to \sqrt{\frac{\theta}{g^2}} \tau \ , \quad
\sigma \to \sqrt{\frac{\theta}{g^2}} \sigma \ , 
\end{eqnarray}
we obtain the effective action
\begin{eqnarray}
S&=&-\frac{\theta}{8 \pi g^2} \int_{-\infty}^\infty d \tau \int_0^{2\pi w} 
d \sigma  \tr \left(
\frac{g^2}{\theta |z|^2}F^2_{z \bar{z}} 
+4 D_+ \phi_i D_- \phi_i 
\right. \n && 
+\frac{\theta |z|^2}{g^2} [\phi_i,\phi_j][\phi_i,\phi_j] 
+2 \bar{\psi} (\Gamma^+ D_+ +\Gamma^- D_-) \psi 
\n && \left.
+2 \sqrt{\frac{\theta}{g^2}}|z| 
\bar{\psi} \Gamma_i [\phi_i,\psi] +O(\theta)
\right) \ . \label{simpleac}
\end{eqnarray}
The parameter $\frac{g^2}{\theta}$ is identified with
the string scale as $\alpha' \equiv \frac{g^2}{\theta}$ in \cite{KN3}.
$\phi_i$ has the target space dimension of $l_s$.
Since the recombination is the local interaction, the length of the
string does not affect the interaction.
Thus, for simplicity, we can consider the two closed strings 
with the equal length.
We parameterize the integral region of $\sigma$ as $[0,2\pi w]$.

This action is valid when two strings coincide in the target space.
If two strings intersect at a point
which is indeed the situation we consider in this paper, 
then the off-diagonal elements
of this action are relevant only
near the intersection point.
They
are meaningful only two strings are very close to 
each other
since otherwise these modes become massive.
Thus, we can describe off-diagonal modes as local fields
$A^{12}_{\tilde{\mu}} 
(\tau,\sigma)$ which are valid near the intersection point with the small
intersection angle.

The definitions of light-cone coordinates 
$\p_{\pm}$ and $A_\pm$ are summarized in 
the appendix \ref{LCC}.
In the free string limit, we obtain light-cone Green-Schwarz 
superstring action which consists of marginal terms. 
In that case, every oscillator mode decomposes into
a left mover and a right mover.

\section{Recombination}
\setcounter{equation}{0}

\subsection{Fluctuation analysis and the recombination}

In this section, by using the multiple closed superstring 
effective action (\ref{simpleac})
obtained in the previous section, we analyze the closed string interactions.
In the IR limit, diagonal components are relevant degrees of freedom.
Thus, the theory becomes a free theory in this limit.
The leading interaction in the perturbative string theory is the interaction
between two closed strings, which is realized as 
the recombination.
The recombination can be investigated by using the Yang-Mills theory 
in \cite{HN}
which is introduced as the low energy effective action of D-strings.

Let us start from the two dimensional effective action
(\ref{simpleac}).
The Bosonic part is written as
\begin{eqnarray}
S=-\frac{\theta}{8 \pi g^2} \int_{-\infty}^\infty d \tau \int_0^{2\pi w} 
d \sigma  \tr \left(
\frac{g^2}{\theta|z|^2}F^2_{z \bar{z}} 
\right. \n
\left.
+4 D_+ \phi_i D_- \phi_i 
+\frac{\theta |z|^2}{g^2} [\phi_i,\phi_j][\phi_i,\phi_j] 
\right) \ . \label{Bos}
\end{eqnarray}
In \cite{KN3}, we identify the overall coefficient $\frac{\theta}{8 \pi
g^2}$ with $\frac{1}{8 \pi \alpha'}$ in the 
free string limit. 
Off-diagonal fields are
regarded as local fields
since we focus on the region where 
two strings are close to each other.

The string coupling is very weak in the IR region.
On the analogy of matrix string theory \cite{DVV},
the string coupling will behave $g_s \propto \frac{1}{|z|}$.
We can interpret this relation as representing the equivalence
between the IR limit and the weak coupling limit.

Supergravity solution of fundamental strings 
which is dual to our effective theory
is described by \cite{IMSY}
\begin{eqnarray}
ds^2 &=& \frac{U^6}{g_{\text{YM}}^4 2^7 \pi^4 N} dx^2 +\frac{1}{2\pi
 g_{\text{YM}}^2} dU^2 +\frac{U^2}{2 \pi g_{\text{YM}}^2} d \Omega^2 \ , \n
e^{\phi} &=&\left(\frac{g^6_{\text{YM}}2^8 \pi^5 N}{U^6} \right)^{-1/2} \ .
\end{eqnarray}
We can read the scaling behavior of $x$ as $x \sim 1/U^3$ from the
metric. 
In the IR limit, the string coupling vanishes $e^{\phi} \sim 
U^3 \sim 1/x \to 0$,
which is consistent with our picture.

It seems that this kind of running coupling behavior makes it difficult 
to treat the interaction. However, the recombination 
happens at a definite scale.
For simplicity, we fix the scale 
of $z$ as $|z|=z_r \equiv 1$ and treat this process in the 
real time\footnote{
For a generic value of $z_r$, the coupling constant is changed
as $g_s^2 \to g_s^2 / z_r^2$.
By taking the following rescaling,
\begin{eqnarray}
q \to \frac{q}{z_r^2} \ , \n
\varphi \to \frac{\varphi}{z_r} \ , \n
\sigma \to \sigma z_r \ ,
\end{eqnarray}
we can absorb this factor.
It is consistent with our claim (\ref{iden}) $q \sim g_s^2$.
}.
We map the coordinates from $z=e^{\tau + i\sigma}$ into 
$z=e^{i(t+\sigma)}$ by the analytic continuation $\tau \to i t$.
Then, the complex plane is mapped into the cylinder 
with the radius $|z|=1$.
The Bosonic part of the action becomes
\begin{eqnarray}
S=-\frac{\theta}{8 \pi g^2} \int_{-\infty}^\infty d t \int_0^{2\pi w} 
d \sigma  \tr \left(
\frac{g^2}{\theta }F^2_{z \bar{z}} \right. \n
\left.
+4 D_+ \phi_i D_- \phi_i 
+\frac{\theta }{g^2} [\phi_i,\phi_j][\phi_i,\phi_j] 
\right) \ . 
\label{2d-IIB}
\end{eqnarray}
This action is the two dimensional SU(2) Yang-Mills theory
in the Lorentzian metric.
Since we are considering the large winding number $w$, the 
worldsheet length 
of the string is very large. Thus, the solution of a single 
closed string might be represented locally as
\begin{eqnarray}
\phi_2&=& \sqrt{\frac{g^2}{\theta}}  \sigma \ , \n
A_\pm&=&\psi=0, \quad \phi_i=0 \ (i=3,\cdots,9) \ ,
\label{sol-t}
\end{eqnarray}
in $U(1)$ gauge theory. 

The solution
which represents the intersecting strings is written as 
\begin{eqnarray}
(\phi_2)_{\text{b.g.}}&=&
\left(
\begin{array}{cc}
q \sqrt{\frac{g^2}{\theta}}
 \sigma & 0 \\
0 &-q \sqrt{\frac{g^2}{\theta}}
 \sigma 
\end{array}
\right)= q \sqrt{\frac{g^2}{\theta}}
 \sigma \sigma^3 \ ,
\n
A_\pm&=&\psi=0, \quad \phi_i=0 \ (i=3,\cdots,9) \ .
\label{solution}
\end{eqnarray}
Two strings intersect at $\sigma=0 $.
Each string is connected with each other at a point
far from the origin $\sigma =0$,
which cannot be seen in this local solution
near the intersection point. 
Thus, if the recombination occurs at this point, 
we obtain the single closed string.
$q$ is a parameter 
which is related to the intersection angle $\phi$ as
\begin{eqnarray}
q \equiv   \tan \frac{\phi}{2} \ .
\label{q-def}
\end{eqnarray}
$q$ is the dimensionful quantity in the worldsheet sense
and as we will see 
later, we relate it 
to the string coupling 
constant $g_s$.
If we take $q \ll 1$,
two strings are very close to each other 
in the target space,
especially around the intersection point.
Thus, our effective theory is valid.

We consider the fluctuations around this background.
We turn on the fluctuations $a_\pm$ and $\varphi$
\begin{eqnarray}
\phi_2=q \sqrt{\frac{g^2}{\theta}}
 \sigma  \sigma^3
+ \sqrt{\frac{g^2}{\theta}}
\varphi \sigma^1, \n
A_+=    a_+ \sigma^2  \ , \quad
A_- =    a_- \sigma^2 \ ,
\label{pa}
\end{eqnarray}
since other fluctuations decouple from these fields
at the quadratic level.
$\sigma^i (i=1,2,3)$ are the Pauli matrices.

The quadratic lagrangian is obtained as
\begin{eqnarray}
L&&=
\frac{g^2}{\theta} \left[
-( \p_+ a_- - \p_- a_+)^2 \right. \n
&&-2( \p_+\varphi -i q a_+  \sigma)
( \p_- \varphi -i  q a_- \sigma)  \n
&&\left.
-  \sqrt{2} q a_- \varphi
+ \sqrt{2} q a_+ \varphi 
\right]
\ . \label{fluc1}
\end{eqnarray}
Note that these fluctuation terms do not come from 
$[\phi_i, \phi_j][\phi_i,\phi_j]$ terms.
By choosing the gauge condition 
\begin{eqnarray}
a_+=-a_- \equiv \frac{a}{\sqrt{2}} \ ,
\end{eqnarray}
the quadratic lagrangian
is written by the parameter $\tau$ and $\sigma$ as
\begin{eqnarray}
L= 
\frac{g^2}{\theta} \left[
 (\p_t a )^2 +  (\p_t \varphi)^2 -
\left(
(\p_\sigma \varphi)^2+2  q \sigma a \p_\sigma \varphi
\right. \right. \n
\left. \left.
+(q  a  \sigma )^2 -2 q a \varphi
\right)  \right]
\ . \label{fluc2}
\end{eqnarray}
In order to solve the equation of motion for the fluctuations, 
we expand the fluctuations by the mass eigenfunctions 
\begin{eqnarray}
a(t,\sigma)=\sum_{n \geq 0}\tilde{a}(\sigma)C_n (t) \ , \n
\varphi (t,\sigma)=\sum_{n \geq 0}\tilde{\varphi}(\sigma)C_n (t) \ ,
\end{eqnarray}
 where $C_n (t)$ satisfy the equations
\begin{eqnarray}
(\p_t^2+m_n^2) C_n(t) =0 \ .
\end{eqnarray}

The differential equations are solved in the appendix \ref{EOM}.
For the lowest mode $n=0$, the eigenfunctions are calculated as
\begin{eqnarray}
C_0 (t) &=&C_0 \cdot \exp (  \sqrt{q }t) \ , \n
\tilde{a}_0 (\sigma)&=&\tilde{\varphi}_0 (\sigma)=\exp 
\left(-\frac{q }{2}  \sigma^2
\right) \ , \label{sol}
\end{eqnarray}
with the eigenvalue
\begin{eqnarray}
m_0^2=-q \ .
\end{eqnarray}
These eigenfunctions are localized near the intersection point
$\sigma   \sim 0$.
This mode is tachyonic and it
causes the recombination \cite{HN}.

\subsection{The recombination probability \label{rec-pro}}

Since the transition probability from two strings into the single string
depends on the string coupling constant $g_s$, 
we can identify $g_s$
by our investigation.
We estimate the recombination probability per unit time.
Since we put two strings with the relative center of mass 
velocity 
$v=0$ as the initial condition,
the recombination always occurs if we wait long enough.

Before calculating the probability, we confirm that the lowest mode
$C_0$ (from now on, we denote it as $C$) indeed causes the recombination.
During and after the recombination, we investigate the time evolution
of the tachyonic mode.
Small fluctuations trigger the recombination and
if we wait long enough, the recombination is over.

The recombination 
can be seen by the diagonalization of 
the background with the off-diagonal fluctuations
\begin{eqnarray}
&&\phi_2 (t,\sigma) 
=
\frac{1}{2}
\sqrt{\frac{g^2}{\theta}}
\left(
\begin{array}{cc}
q   \sigma 
&  C(t) \tilde{\varphi}_0 (\sigma) \\
 C (t) \tilde{\varphi}_0 (\sigma) &-  q \sigma
\end{array}
\right) \n 
&& \hspace*{1cm}
\xrightarrow{\text{diag}}
\frac{1}{2}
\sqrt{\frac{g^2}{\theta}} \times \n
&&\left(
\begin{array}{cc}
\sqrt{(q  \sigma )^2
+C^2 (t)e^{-q  \sigma^2}}& 0 \\
0 &-\sqrt{(q 
  \sigma )^2+C^2 (t)e^{- q
\sigma^2}}
\end{array}
\right) \ . \n
\label{recombi}
\end{eqnarray}

Note that although this phenomenon locally represents 
the recombination of 2 strings $\to$ 2 strings, 
the final state
is connected globally.
Thus, this recombination represents the transition
from two closed strings into the single closed string.
We need to clarify the
validity of the approximation since
this geometrical picture comes 
from the fluctuation analysis.
Our analysis is valid if
the terms quadratic in the fluctuations are much smaller
than the terms in the background.
This condition is 
\begin{eqnarray}
(q  \sigma)^2 \gg C^2 e^{-q \sigma^2} \ .
\end{eqnarray}
Since $q \sigma^2 \sim O (1)$
as can be seen in (\ref{inv}),
we obtain 
\begin{eqnarray}
q  \gg C^2 \ .
\end{eqnarray}
Our analysis is valid under this condition.

The action quadratic in the $C $ mode
is obtained as
\begin{eqnarray}
S=\frac{1}{8 \pi } \int_{-\infty}^\infty dt \int_{-\pi w}^{\pi w} 
d\sigma
\left( (\p_t C (t)
)^2 +q C^2 (t) \right) 
\n \times 
\exp \left( -\frac{q}{2}
 \sigma^2 \right) \n
\sim \frac{1}{8 \pi } \sqrt{\frac{\pi }{2  q}}
\int_{-\infty}^\infty dt
\left( (\p_t C (t))^2 +q C^2 (t)\right) \ .
\hspace*{0.2cm}
\label{inv}
\end{eqnarray}
After integrating the $\sigma$ direction,
this action can be regarded as a quantum mechanics of a particle 
moving in the inverse harmonic oscillator \cite{HH}.
As seen in (\ref{recombi}),
$C (t)$ is related to the separation of recombined strings.
If $C$ takes a large value, then we can regard it as the signal for
recombination. 
Once $C$ obtains the large
value, the recombination is over.
The separation of strings grows 
and finally
we obtain a single closed string.
In the inverse harmonic oscillator potential, the wave functions of the 
particle will keep spreading to the larger $|C|$.
Thus, the recombination probability approaches 1 at $t \to \infty$.

The parameters in (\ref{inv}) can be interpreted in terms of the 
quantum mechanics
of the particle moving in the inverse harmonic potential as
\begin{eqnarray}
m &\equiv& \frac{1}{8 \pi } \sqrt{\frac{\pi }{2 q}} \ , \n
\omega &\equiv& \sqrt{q } \ , 
\end{eqnarray}
where $m$ is the mass of the particle and $\omega$ is the frequency. 

The Schrodinger equation is given by
\begin{eqnarray}
i \frac{\p \psi}{\p t}=-\frac{1}{4m}\frac{\p^2 \psi}{\p C^2 }
-m\omega^2 C^2 \psi \ . \label{Sch}
\end{eqnarray}
As an initial condition at $t=0$,
we consider the wave function to be a gaussian
which is labeled by the parameter $\phi$.
The derivation of the solution of the equation (\ref{Sch})
is discussed in the appendix \ref{SCH2}.
For large $t$, the wave function behaves
\begin{eqnarray}
\psi (C, t) \sim (2/\pi )^{1/4} b^{-1/2} \exp \left(
-\frac{1}{2}(\omega t + i \phi)
\right)
\n
\times
\exp \left(
-e^{-2\omega t } \frac{C^2}{b^2}+i m\omega C^2
\right) \ . \label{wave}
\end{eqnarray}
As discussed previously, 
once recombination happens, we obtain the single closed string as a
final state.
Since the fluctuation analysis is valid in the region
$\sqrt{q} 
\gg C$, the geometric picture is also reliable in this region.
We judge that the recombination has happened if the value of $C (t)$ 
grows beyond $\sqrt{q}=\omega$.
Thus, the recombination probability at a time $t$ is
estimated as
\begin{eqnarray}
P(t) &=&2\int_{\omega}^\infty 
d C |\psi (C,t)|^2 \n
&=&1-\text{Erf}( \sqrt{2}b^{-1} e^{-t \omega} 
\omega ) \n
&=&1-\text{Erf} \left(\sqrt{
4 m \omega^3
          \sin 2 \phi}  e^{-t \omega} \right) \ .
\end{eqnarray}
One can confirm that
at $t \to \infty$, $P \to 1$.
The recombination probability per unit time is calculated as
\begin{eqnarray}
\frac{d P}{d t}&=&\frac{2}{\sqrt{\pi}} \sqrt{4 m \omega^5 \sin 2 \phi}
\exp \left(
-4 m \omega^3 \sin 2 \phi e^{-2 t \omega}-t \omega
\right) \n
&=&\frac{2^{1/4}}{\pi^{3/4}} q \sqrt{\sin 2 \phi}
\n
&&\times
\exp
\left(
-\frac{q}{2 \sqrt{2 \pi}} \sin 2 \phi e^{-2 \sqrt{q}t}-\sqrt{q}t
\right)
\ . \label{prob-IIB}
\end{eqnarray}
In the small $q$ limit,
the probability is proportional to 
\begin{eqnarray}
\frac{dP}{dt} \propto q \ .
\end{eqnarray}
In the perturbative string calculation, this probability 
is proportional to $g_s^2$.
Thus, we identify
\begin{eqnarray}
q \sim g_s^2 \ .
\label{iden}
\end{eqnarray}
The higher order corrections come from the expansion of the exponential
function. By expanding this function
at a time $t \sim O(\frac{1}{\sqrt{q}})$,
and in the small $q$ limit, 
we obtain 
\begin{eqnarray}
\sum_{n=1}^\infty P_n q^{n} \ , 
\end{eqnarray}
which is consistent with the higher order corrections of the 
perturbative string if we identify $q$ with $g_s^2$
\cite{Polchinski,JJP}.

\subsection{The recombination in matrix string theory}

In this section, we estimate the recombination probability of
intersecting strings from matrix string theory.
The action of matrix string theory is written 
by the two dimensional Yang-Mills theory 
\begin{eqnarray}
S&=& \frac{1}{l_s^2}\int_{-\infty}^\infty d \tau \int_0^{2\pi w} 
d \sigma  \tr \left(
(l_s^2 g_s^2)
F^2_{z \bar{z}} 
\right. \n
&&+2 (D^+ \phi_i D_+ \phi_i +D^- \phi_i D_- \phi_i)
 \n && 
+\frac{1}{l_s^2g_s^2} [\phi_i,\phi_j][\phi_i,\phi_j] 
+2 \bar{\psi} (\Gamma^+ D_+ +\Gamma^- D_-) \psi 
\n
&&+\left. \frac{2}{g_sl_s} \bar{\psi} \Gamma_i [\phi_i,\psi] 
\right) \ ,
\label{MST}
\end{eqnarray}
where we consider long intersecting strings with the large
winding number $w$.
By the identification $\frac{\theta}{g^2} \equiv \frac{1}{l_s^2}$,
the action (\ref{MST}) is very close to the action (\ref{2d-IIB}).
The different point is the $g_s$ dependence which appears explicitly
in (\ref{MST}).
Thus, we perform the fluctuation analysis for the action (\ref{MST})
and investigate the $g_s$ dependence.
Since the action is the same apart from the $g_s$ dependence, 
if we consider the same solution, 
we obtain the same tachyon mode 
as in section 3.1.

By regarding the tachyon effective action as the quantum mechanics 
of the particle moving in the inverse harmonic oscillator,
the parameters $g_s$ and $q$ appear in the mass and frequency of the
particle as 
\begin{eqnarray}
m &\equiv& \frac{g_s^2}{8 \pi} \sqrt{\frac{\pi}{2 q}}
\ , \n
\omega &\equiv& \sqrt{q} \ .
\end{eqnarray}
The only different point with respect to the analysis in 
section 3.1 is the $g_s$ dependence.
The recombination probability at a time $t$ is estimated 
from the previous calculation (\ref{prob-IIB}) as
\begin{eqnarray}
P(t) =1- \text{Erf} \left(
\sqrt{4g_s^2 \sqrt{\frac{\pi}{2}} q \sin 2 \phi}
e^{-\sqrt{q}t}
\right) \ .
\end{eqnarray}
Thus, the recombination probability per unit time is
\begin{eqnarray}
\frac{dP}{d t} &=&\frac{2^{1/4}}{\pi^{3/4}} g_s q\sqrt{\sin 2 \phi}
\n
&\times&
\exp \left(
-\frac{1}{2\sqrt{2\pi}} g_s^2 q \sin 2 \phi e^{-2 t
\sqrt{q}}
-\sqrt{q}t
\right)
\ .\hspace*{1cm}
\end{eqnarray}
By the rescaling
\begin{eqnarray}
t \to g_s t \ ,
\end{eqnarray}
we obtain
\begin{eqnarray}
\frac{dP}{d t} &=&\frac{2^{1/4}}{\pi^{3/4}} g_s^2 q\sqrt{\sin 2 \phi}
\n
&\times& \exp \left(
-\frac{g_s^2 q}{2\sqrt{2\pi}} \sin 2 \phi e^{-2 g_s t
\sqrt{q}}
-\sqrt{q} g_s t
\right)
\ .\hspace*{1cm}
\end{eqnarray}
This result is equivalent with the previous result (\ref{prob-IIB})
if we replace $q \leftrightarrow q g_s^2$.
Thus, the identification we have derived in the section
\ref{rec-pro} is consistent with matrix string theory.

At a large time $t \sim O(\frac{1}{g_s\sqrt{q}})$
in the small $q$ limit, the leading contribution is 
proportional to $g_s^2 q$.
The higher order corrections which come from the expansion of the 
exponential part are 
\begin{eqnarray}
(\text{leading contribution}) \times \sum_{n=1}^\infty P_n g_s^{2n} \ ,
\end{eqnarray}
which is consistent with the perturbative string theory.

In \cite{BBN}, the classical BPS solutions that interpolate between 
the initial and the final string configurations are constructed in 
matrix string theory.
They interpret the amplitude of matrix string theory as 
the transition amplitude between initial and final configurations,
and show that the leading contribution is proportional to 
$g_s^{- \chi}$ where 
$\chi$ is the Euler characteristic of the interpolating Riemann
surface.
Thus, 
they have reproduced the perturbative string amplitude
from matrix string theory,
which is consistent with our result.

\section{Conclusion}
\setcounter{equation}{0}

We have identified the string coupling $g_s$ in IIB matrix model.
We have constructed the classical solution of the strings in the action
which is obtained from IIB matrix model.
Starting from the configuration with the intersection angle $\phi =2
\tan^{-1} q$ and no transverse distance and relative velocity, 
the recombination happens.
This is triggered by the tachyonic fluctuations around the classical
solution. 
After the recombination, we obtain the single closed string.
We have estimated the probability of the recombination of two strings
per unit time in this situation.
The recombination probability is 
also calculated by the perturbative string theory.
The leading contribution 
is proportional to 
$g_s^2$. The higher order corrections are seen as $\sum_{n=2}^\infty
P_n g_s^{2n}$. Comparing our results with these behaviors,
we have identified the string coupling $q \sim g_s^2$.

We have also estimated the recombination probability per unit time in
the matrix string theory.
The result shows the consistent behavior with that of the 
perturbative string theory.
In the matrix string theory,
the Yang-Mills coupling has the dimension $[length]^{-1}$ in the
worldsheet and it is inversely proportional to the string coupling $g_s$.
Thus, $g_s$ has the dimension $[length]$ in the worldsheet.
Since $q$ has the worldsheet dimension $[length]^{-2}$, the
dimensionless parameter is $q g_s^2$.
Since the probability is the dimensionless quantity, 
$q$ should appear with $g_s^2$
which is consistent with the result obtained from IIB matrix model.

The parameter $q$ represents
the ratio between the target space coordinate $\phi_i$ 
and the worldsheet coordinate $\sigma$.
Another parameter which has this kind of property
is the velocity $v$.
Although the recombination probability depends on the relative velocity 
$v$,
we have put $v=0$ in this paper.
It might be interesting to calculate the probability of the
recombination of intersecting strings with nonzero relative velocity.

\begin{center} \begin{large}
Acknowledgments
\end{large} \end{center}
This work is supported in part by the Grant-in-Aid for Scientific
Research from the Ministry of Education, Science and Culture of Japan.

\appendix

\section{Notation}
\setcounter{equation}{0}

\subsection{Light-cone coordinates \label{LCC}}

$\p_\pm$ are the derivatives with respect to 
\begin{eqnarray}
y_\pm \equiv \frac{1}{\sqrt{2}}(\tau \pm i \sigma) \ ,
\end{eqnarray}
and they are related to $\p_0$ and $\p_1$ as
\begin{eqnarray}
\p_0 =\frac{1}{\sqrt{2}}(\frac{\p_+}{z}+\frac{\p_-}{\bar{z}}) \ , \quad
\p_1= \frac{i}{\sqrt{2}} (\frac{\p_+}{z}-\frac{\p_-}{\bar{z}} ) .
\end{eqnarray}
$A_\pm$ are related to $A_0$ and $A_1$ as
\begin{eqnarray}
A_0 =\frac{1}{\sqrt{2}}(A_+ + A_-) \ , \quad
A_1= \frac{i}{\sqrt{2}}(A_+ - A_-) \ .
\end{eqnarray}

The definition of $F_{z\bar{z}}$ in (\ref{simpleac}) is 
\begin{eqnarray}
F_{z\bar{z}} &\equiv& \bar{z}\p_+ (\bar{z}^{-1} A_- )
- z\p_- (z^{-1}A_+ )-[A_+,A_-] \n
&=& \p_+ A_- -\p_- A_+ -[A_+,A_-]
\ .
\end{eqnarray}

\subsection{Pauli matrices \label{Pauli}
}

The Pauli matrices in (\ref{pa}) satisfy the following relation
\begin{eqnarray}
[\sigma^i, \sigma^j]=i \epsilon^{ijk}\sigma^k \ , \quad
\tr (\sigma^i)^2  =\frac{1}{2} \ .
\end{eqnarray}

\section{Calculation}
\setcounter{equation}{0}

\subsection{The equation of motion for (\ref{fluc2})
\label{EOM}}

The derivation of the eigenfunctions (\ref{sol})
is summarized in this appendix.

By using the relation
\begin{eqnarray}
\p_+ +\p_- &=&-\sqrt{2}i \p_t \ , \n 
2 \p_+ \p_- &=&-\p_t^2 +\p_\sigma^2 \ , \n
\p_+-\p_- &=&- \sqrt{2}i\p_\sigma \ ,
\end{eqnarray}
and imposing
the gauge fixing condition $a_+ =-a_- \equiv \frac{a}{\sqrt{2}}$, we obtain 
the lagrangian (\ref{fluc2}).

The equation of motion for the fluctuation lagrangian (\ref{fluc2})
is 
\begin{eqnarray}
\hat{O} 
\left(
\begin{array}{c}
a(t,\sigma) \\
\varphi (t,\sigma) 
\end{array}
\right)=0 \ ,
\end{eqnarray}
where
\begin{eqnarray}
\hat{O}=
\left(
\begin{array}{cc}
-(q \sigma )^2 - \p_t^2& - q 
 \sigma \p_\sigma +q \\
q \sigma \p_\sigma+2 q & -\p_t^2 +\p_\sigma^2 
\end{array}
\right) \ .
\end{eqnarray}
For the mass eigenvalues $m_n^2$
which satisfy the free field equation
\begin{eqnarray}
(\p_t^2 +m_n^2) C_n (\tau) =0 \ ,
\end{eqnarray}
the equation of motion is given by
\begin{eqnarray}
\left(
\begin{array}{cc}
-(q \sigma)^2 + m_n^2& - q \sigma \p_\sigma 
+q \\
q \sigma \p_\sigma+2 q & \p_\sigma^2 +m_n^2 
\end{array}
\right)
\left(
\begin{array}{c}
\tilde{a}(\sigma) \\
\tilde{\varphi} (\sigma) 
\end{array}
\right)=0 \ .
\end{eqnarray}
This differential equation is solved with the mass eigenvalue
\begin{eqnarray}
m_n^2=(2n-1) q \ .
\end{eqnarray}
For the lowest mode $n=0$, the eigenfunctions are calculated as
\begin{eqnarray}
C_0 (\tau) &=&C_0 \cdot \exp (  \sqrt{q }t) \ , \n
\tilde{a}_0 (\sigma)&=&\tilde{\varphi}_0 (\sigma)=\exp 
\left(-\frac{q }{2 }  \sigma^2
\right) \ .
\end{eqnarray}
For general $n$, the eigenfunctions are 
\begin{eqnarray} 
\tilde{a}_n 
(\sigma)=&&- e^{-\frac{q\sigma^2}{2}} \sum_{j=0,2,\cdots}^n (-1)^{\frac{j}{2}} 
\frac{4^{\frac{j}{2}}}{j!}
\frac{n(n-2)\cdots (n-j+2)}{2n-1} 
\n
&&\times (j-1) 
\left( \sigma \sqrt{\frac{q}{2}}\right)^j \ , \n
\tilde{\varphi}_n 
(\sigma)=&& e^{-\frac{q\sigma^2}{2}}\sum_{j=0,2,\cdots}^n (-1)^{\frac{j}{2}} 
\frac{4^{\frac{j}{2}}}{j!}
\frac{n(n-2)\cdots (n-j+2)}{2n-1} \n
&& \times (2n\!-\!j\!-\!1) 
\left(\sigma \sqrt{\frac{q}{2}} \right)^j 
\end{eqnarray}
for $n=0,2,\cdots$, and
\begin{eqnarray}
\tilde{a}_n (\sigma)=&&
- e^{-\frac{q\sigma^2}{2}}\sum_{j=1,3,\cdots}^n (-1)^{\frac{(j-1)}{2}}
 \frac{4^{\frac{(j-1)}{2}}}{j!}
\left(\frac{j-1}{2}\right) 
\n &&\times (n-3) \cdots (n-j+2)
\left(\sigma \sqrt{\frac{q}{2}} \right)^j \ , \n
\tilde{\varphi}_n (\sigma)=&&
 e^{-\frac{q\sigma^2}{2}}\sum_{j=1,3,\cdots}^n (-1)^{\frac{(j-1)}{2}} 
\frac{4^{\frac{(j-1)}{2}}}{j!}
\left(n-\frac{j+1}{2}\right) \n
&&\times (n-3) \cdots (n-j+2)
\left(\sigma \sqrt{\frac{q}{2}} \right)^j  
\end{eqnarray}
for $n=3,5,\cdots$. 

\subsection{Solving the Schrodinger equation (\ref{Sch})
\label{SCH2}}

We summarize the derivation of the solution (\ref{wave}) in this appendix.

We describe the momentum conjugate to the $C$ as $\Pi$
\begin{eqnarray}
\Pi =\frac{\delta L}{\delta \dot{C}}= 2 m \dot{C} \ .
\end{eqnarray}
Hamiltonian is given by
\begin{eqnarray}
H&=&\Pi \dot{C}-L \n
&=&\frac{1}{4m} \Pi_0^2 -m\omega^2 C_0^2 \ .
\end{eqnarray}

The Schrodinger equation (\ref{Sch}) is solved 
by the wave function of the form 
\begin{eqnarray}
\psi (C,t) =A(t) \exp (-B(t) C^2) \ ,
\end{eqnarray}
if $A$ and $B$ satisfy the equation
\begin{eqnarray}
i \dot{A}&=&\frac{1}{2m}AB \ , \n
\frac{i}{m}  \dot{B}&=&\omega^2 +\frac{1}{m^2} B^2 \ .
\end{eqnarray}
These are solved in \cite{GP} as
\begin{eqnarray}
A&=&(2 \pi)^{-1/4} \left( b \cos (\phi -i \omega t)\right)^{-1/2} \ ,
 \n
B&=& m\omega \tan (\phi - i \omega t )\n
&=& m \omega \frac{\sin 2 \phi -i \sinh 2 \omega t}{\cos
2 \phi +\cosh 2 \omega t} \ ,
\end{eqnarray}
where
\begin{eqnarray}
a^2 &\equiv& \frac{1}{2 m\omega}
 \ , \n
b&=&a (\sin 2\phi)^{-1/2} \ .
\end{eqnarray}
We put the initial condition of the wave function as
\begin{eqnarray}
\psi (C, t=0) =(2\pi)^{-1/4} 
(b\cos \phi)^{-1/2} 
\n \times
\exp
(-m \omega \tan \phi C^2) \ . 
\end{eqnarray}
The parameter $\phi$ controls the initial condition.

\end{document}